\documentclass[a4paper,12pt]{article}
\usepackage{graphicx}
\usepackage{graphics}

\newcommand{\dfr}{\bar{d}}
\newcommand{\Csc}{I \!\!\!\! C}

\newcommand{\Planck}{I \!\!\!\! P}

\def\d{\partial}

\begin{document}

\title{Non-Abelian gauge field theory in scale relativity}
\author{Laurent Nottale\footnote{Electronic mail: laurent.nottale@obspm.fr},   
Marie-No\"elle C\'el\'erier\footnote{Electronic mail: marie-noelle.celerier@obspm.fr}  
and Thierry Lehner\footnote{Electronic mail: thierry.lehner@obspm.fr}\\
 {\it  \small Observatoire de Paris-Meudon, LUTH, CNRS,
5 place Jules Janssen, 92195 Meudon Cedex, France}}
\maketitle

\abstract

Gauge field theory is developed in the framework of scale relativity. In this theory, space-time is described as a nondifferentiable continuum, which implies it is fractal, i.e., explicitly dependent on internal scale variables. Owing to the principle of relativity 
that has been extended to scales, these scale variables can themselves become  functions of the 
space-time coordinates. Therefore, a coupling is expected between displacements 
in the fractal space-time and the transformations of these scale variables. 
In previous works, an Abelian gauge theory (electromagnetism) has been derived 
as a consequence of this coupling for global dilations and/or contractions. 
We consider here more general transformations of the scale 
variables by taking into account separate dilations for each of them, which yield non-Abelian gauge theories. We identify these 
transformations with the usual gauge transformations. The gauge fields naturally 
appear as a new geometric contribution to the total variation of the action 
involving these scale variables, while the gauge charges emerge as 
the generators of the scale transformation group. A generalized action is 
identified with the scale-relativistic invariant. The gauge charges are 
the conservative quantities, conjugates of the scale variables through the
action, which find their origin in the symmetries of the ``scale-space''.
We thus found in a geometric way and recover the expression for the covariant derivative of gauge 
theory. Adding the requirement that under the scale transformations the fermion
multiplets and the boson fields transform such that the derived 
Lagrangian remains invariant, we obtain gauge theories as a 
consequence of scale symmetries issued from a geometric space-time description.

%1***************

\section{Introduction}
\label{s.i}

%****************

In standard gauge field theory, the nature of the gauge transformations, of
the gauge fields and of the conserved charges are postulated and 
designed from experimental considerations. The group of gauge
transformations does not act upon the space-time coordinates, as does, for
example, the SU(2) spin rotation group or the Lorentz group, but in an
``internal space'' whose physical meaning is not understood from first principles. For a general
gauge group G, the particle wave functions that are multiplets of Dirac 
bi-spinors form a $n$-component
vector in the internal space, and the gauge potentials $A_{\mu }$ (more
generally $W_{\mu }^{a}$) are fields in standard space-time defined only up
to a gauge transformation.

There is indeed a fundamental difference between the situation of
transformations in the standard gauge theories and of, e.g., Lorentz
transformations. Thanks to the fact that space-time coordinates are directly observable, we know from the very beginning what Lorentz
transformations are, namely, space-time rotations of the coordinates, $dx^{\prime }=\Lambda _{\beta }^{\alpha } \, dx^{\beta }$. They write 
in the case of an infinitesimal transformation, (i) $dx^{\prime }{}^{\alpha
}=(\delta _{\beta }^{\alpha }+\omega _{\beta }^{\alpha })dx^{\beta }$, where the $\omega^{ij }$ ($i$ and $j=1$ to $3$) represent the infinitesimal angles of rotation in space and the $\omega^{0i}=v^i/c$ ($v^i \ll c$) are the infinitesimal Lorentz boosts. Then, once this basic
definition is given, one can consider the effect of these transformations on
various physical quantities defined in space-time, e.g., the wave function  $\psi $. This involves representations of the Lorentz group adapted to the nature of the physical
object under consideration, i.e., (ii) $\psi ^{\prime }=(1+\frac{1}{2}\omega
^{\alpha \beta }\sigma _{\alpha \beta })\psi $ (see, e.g.,  \cite{SW72}). 

In contradistinction with this situation, in standard gauge theories the gauge functions, being arbitrary, are considered to be devoid of physical meaning. As a consequence,  there is up to now no equivalent of the basic defining transformation (i). Therefore, in the standard framework, the gauge group is indirectly defined through its action on the various physical
objects according to its representations, in similarity with relation (ii), but the physical meaning of the gauge space itself is lacking.

In the present paper, we place ourselves in the framework of the scale relativity
theory, in which the description of the space-time geometry is generalized to continuous but nondifferentiable `manifolds'. In this theory, one attempts to recover the quantum behavior as a manifestation of the nondifferentiability, then the gauge fields themselves as a manifestation of  the nondifferentiable and fractal geometry (in analogy with gravitation interpreted as a manifestation of the non-Euclidean curved geometry in general relativity). 

 In this framework, we give a geometric meaning to the gauge space,  then we can rebuild the gauge transformations of the various physical quantities (namely, the various quantum fields) as consequences of the fundamental transformations of the variables which define this gauge space.  In other words, it is precisely an equivalent for gauge theories of the defining transformation (i) that can be proposed in scale relativity. The specifically new results given in the present  paper consist of extending to non-Abelian gauge theory the results of previous works \cite{LN94,LN96,LN03} devoted to the understanding of the simpler gauge invariant theory of electromagnetism.

The paper is organized as follows. After a summarized review (Sec. \ref{s.de}) of the main steps of the construction of the scale relativity theory, including the exposition 
of the salient features that have led to the demonstration of the Dirac 
equation \cite{CN03,CN04} from the scale relativistic 
first principles, we give a brief reminder of the results 
previously obtained for electromagnetism
(Sec. \ref{s.elec}). Then we give, in Sec. \ref{s.nagf}, an extension of the concepts
and methods thus obtained, and we apply them to a general development of the
non-Abelian gauge formalism. Section \ref{s.c} is devoted to the conclusion.

%2**********************

\section{Scale relativity and quantum mechanics: summary}
\label{s.de}

%********************************

%**2.1***********************************
\subsection{Foundations of scale relativity}

The theory of  scale relativity is based on 
the giving up of the hypothesis of manifold differentiability which is a key 
assumption of Einstein's general relativity. In the new theory, the 
coordinate transformations are continuous but can be differentiable (and 
therefore it includes general relativity) or nondifferentiable. The giving up of the 
assumption of differentiability implies several consequences 
\cite {LN93}, leading to the successive steps of the construction of the theory:

(1) It has been proved \cite{LN93,LN96,BAC00} that a continuous and nondifferentiable curve 
is fractal in a general meaning, namely, its length is explicitly
scale dependent and goes to infinity when the scale
interval $\varepsilon$ goes to zero, i.e., ${\cal L} ={\cal L}(\varepsilon) \rightarrow \infty$ when ${\varepsilon \rightarrow 0}$. This result can be readily extended to a continuous and nondifferentiable manifold.
 
 (2) The fractality of space-time \cite{GO83,NS84} involves the scale 
dependence of the reference frames. We therefore add to the 
usual variables defining the reference frames (position, orientation, motion), new 
variables $\varepsilon$ characterizing their `state of scale'. In particular, the coordinates themselves become functions of these scale variables, i.e., $X=X(\varepsilon)$ (in the simplified case of only one variable). In an experimental situation, these scale variables are identified with the resolution scale of the measurement apparatus. In the case of a theoretical physics description, they are identified with the differential elements themselves, of which the coordinates become explicit functions, i.e., $X=X(dX)$.
 
 (3) The scale variables $\varepsilon$ can never be defined in an absolute way, but only in a relative way. Namely, only their ratio $\rho=\varepsilon'/\varepsilon$ does have a physical meaning. This universal behavior leads to extend to scales the principle of relativity \cite{LN89,LN92,LN93}, in order to include in the possible changes of reference frames the new ones which are described by the transformations of these scale variables.

(4) Though the nondifferentiability manifests itself at the limit $\varepsilon \rightarrow 0$,  the use of differential equations is made 
possible by representing physical quantities $f$ by fractal functions $f[X(\varepsilon), \varepsilon]$ \cite{LN93}. Even if the function $f(X,0)$ is nondifferentiable with respect to the variable $X$, the fractal 
function $f(X, \varepsilon)$ is differentiable for any $\varepsilon \neq 0$ 
with respect to both $X$ and $\varepsilon$. This allows us to complete the 
differential 
equations of standard physics by new differential equations of scale, which are constrained by the principle of scale relativity. The study of the scale laws derived from these differential equations has been developed according to various levels of relativistic transformations \cite{LN92,LN96,LN97a}. In what follows, we consider only the simplest case, namely Galilean-type scale transformations (i.e., characterized by a constant fractal dimension).
  
 (5) The simplest possible scale differential equation is a first order equation, $\partial X / \partial \ln \varepsilon=\beta(X)$, which can be simplified again by Taylor expanding the unknown function $\beta$, so that it reads ${\partial X} / {\partial \ln \varepsilon}= a+ b X + \cdots$.
 The solution of this equation is made of two terms, a scale-independent, differentiable, classical part and a power-law, nondifferentiable fractal part, which read
 \begin{equation}
 \label{equ2}
X=x +\zeta \left( \frac{\lambda}{\varepsilon} \right)^{-b},
\end{equation}
 where $x=-a/b$. When the coefficient $b$ is constant, the second term is the standard expression for the length of a fractal curve of dimension $D_{F}=1-b$ \cite{BM82}. Moreover, the transformation law of this power-law term under a scale transformation $ \ln(\lambda/\varepsilon) \rightarrow \ln(\lambda/ \varepsilon')$ takes the mathematical form of the Galileo group, and it therefore comes under the principle of relativity \cite{LN92}, as initially required.
 
% 2.2 **********************
\subsection{Metric of a fractal space-time}
\label{sec:metric}
%**********************
  In Eq.~(\ref{equ2}), the scale variable $\varepsilon$ is a  space resolution, e.g., $\varepsilon=\delta X$. The next step consists of  considering  its four-dimensional differential counterpart and to express it in terms of intervals of the invariant length (proper time) $ds$, by using the standard relation between the resolution 
interval of projected coordinates and the resolution interval of the invariant length on a fractal, $(\delta X^{\mu})^{D_F}\sim \delta s$,
  \begin{equation}
  \label{equ3}
dX^{\mu} = dx^{\mu} + d\xi^{\mu}= v^{\mu} ds + \zeta^{\mu}  \times (\lambda_c)^{1-1/D_F } \times
ds^{1/D_F},
\end{equation}
where $\zeta^{\mu}$ is dimensionless, $\lambda_c$ is a length scale which must be introduced for dimensional reasons and $D_F$ is a fractal (covering) dimension. In the case where this description holds for a quantum particle of mass $m$, $\lambda_c$ will be identified with its Compton length $\hbar/mc$. The elementary displacement on a fractal space-time is therefore the sum of a classical, standard differentiable element, which is leading at large scales, and of a fractal, nonstandard fluctuation which is leading at small scales. 
 
 In what follows, we simplify again the description by considering only the case $D_{F}=2$. For this, we base ourselves on Feynman's result  \cite{RF48,FH65} according to which the typical paths of quantum particles (those which contribute mainly to the path integral) are nondifferentiable and (in modern words) fractal of dimension $D_{F}=2$. The  case $D_{F}\neq 2$ has also been studied in detail: it has been shown that $D_{F}=2$ is a critical dimension for which the explicit scale dependence disappears in the final equations  (see  \cite{LN96} and references therein).
 
Let us now show how Eq.~(\ref{equ3}) can be used to give an explicit form to the metric of a fractal space-time (disregarding at this step of the construction other consequences of nondifferentiability such as the multivaluedness of derivatives, see next sections). The fractal fluctuations (here in four dimensions) write for fractal dimension 2,
\begin{equation}
\label{0125}
d \xi^{\mu}=\zeta^{\mu} \sqrt{\lambda_c \, ds},
\end{equation}
where the $\zeta^{\mu}$ are dimensionless highly fluctuating functions. 

In what follows, we replace them (in a provisional way) by stochastic variables such that $\langle \zeta^{\mu} \rangle=0$, $\langle(\zeta^0)^2\rangle=-1$ and $\langle(\zeta^k)^2\rangle=1$ ($k=$1 to 3). We recover here a description which is familiar in usual stochastic processes, which can also be separated in a regular part and a stochastic part, but here this is done at the level of the metric. As we shall see, we do not have to be more specific about the probability distribution of these stochastic variables. Their zero mean and unit variance is the only information needed in the subsequent calculations, which are therefore valid whatever this distribution.

Now we can write the fractal fluctuations in terms of the coordinate differentials instead of the invariant length differential,
\begin{equation}
\label{0126}
d \xi^{\mu}=\zeta^{\mu} \sqrt{\lambda^{\mu} \, dx^{\mu}}.
\end{equation}
The identification of Eqs. (\ref{0125}) and (\ref{0126}) leads very simply to the establishment of the expressions for the de Broglie-Einstein length and time scales from the Compton one, i.e., for two variables,
\begin{equation}
\lambda_{x}=\frac{\lambda_c}{dx/ds}=\frac{\hbar}{p_x}, \;\;\;\tau=\frac{\lambda_c}{dt/ds}=\frac{\hbar}{E}.
\end{equation}
%added in revised version
The de Broglie scale (and the Compton scale in rest frame) therefore plays an essential role in the properties of the scale variables  (identified here with the differential elements). It  stands out as a natural reference scale for them, since it plays the role of a fractal to nonfractal transition (that should not be understood as a transition acting in position space but instead  in scale space). Indeed we see from the relation $\langle d \xi_x^2 \rangle=\lambda_x dx$ (and similar relations for the other variables) that when $|dx| \ll \lambda_x$, the fractal fluctuation becomes $|d \xi_x| \gg |dx|$ and therefore it dominates the classical (differentiable) contribution. On the contrary, when $|dx| \gg \lambda_x$, the fractal fluctuation $|d \xi_x| \ll |dx|$ becomes negligible and only the classical term remains. The subsequent developments of the theory, which lead to construct a wave function and to derive Schr\"odinger and Dirac equations (see Sec.~\ref{sec:dirac}), finally allow one to identify this transition with a quantum to classical transition \cite{LN93}.
%end of added text

Let us now assume that the large scale (classical) behavior is given by Riemannian metric potentials ${g}_{\mu \nu}(x,y,z,t)$. The invariant proper time $dS$ along a geodesic (which is therefore subjected to curvature at large scale and fractality at small scales) writes in terms of the complete differential elements $dX^{\mu}=dx^{\mu}+d \xi^{\mu}$: 
\begin{equation}
d{S}^2={g}_{\mu \nu} dX^{\mu} dX^{\nu}={g}_{\mu \nu} (dx^{\mu}+d \xi^{\mu}) (dx^{\nu}+d \xi^{\nu}).
\end{equation}
Now replacing the $d \xi$'s by their expression (Eq.~\ref{0126}), we obtain a fractal metric. Assuming for simplicity (1+1) dimensions, a diagonal classical part of the metric and a fractal dimension $D_F=2$, it reads
\begin{equation}
d{S}^2={g}_{00} \left( 1+ \zeta_{0}\; \sqrt{ \frac{\tau}{dt}}  \right)^2  c^2 dt^2-{g}_{11} \left( 1+ \zeta_{1}\; \sqrt{ \frac{\lambda_x}{dx}}  \right)^2 dx^2.
\end{equation}
We therefore obtain generalized fractal metric potentials which are explicitly dependent on the coordinate differential elements, in agreement with the program of Refs.~\cite{LN89,LN93}. More generally the metric potentials can be written in their turn as the sum of the standard metric potentials (which describe curvature) and of divergent, highly fluctuating terms (which describe fractality), e.g., for the $g_{00}$ component,
\begin{equation}
\tilde{g}_{00}(x,t;dt)=g_{00}(x,t)+\gamma_{00}(x,t) \left( \frac{\tau}{dt}  \right),
\end{equation}
where we have kept only the leading term, owing to the fact that $\langle \zeta^{\mu} \rangle=0$.  The $\gamma_{\mu \nu}(x,t)$ can be described at a first approximation in terms of stochastic variables. We recover here our result \cite{LN93} according to which, in the limit ($dx,\,dt\rightarrow0$), the metric is divergent (singular) at each of its points and instants,  which is the very intrinsic expression of the fractality of space-time. As a consequence, the curvature is also explicitly scale-dependent and divergent when the scale intervals tend to zero. This property ensures the fundamentally non-Riemannian character of a fractal space-time as well as the ability to characterize it in an intrinsic way.

%added in revised version
Note that all the above developments have been made in the framework of Galilean scale relativity, in which the fractal dimension is assumed to be constant (we call it Galilean because its laws of scale transformation are similar to inertial laws of motion). However it is worth briefly recalling here that a special scale relativity theory has been proposed \cite{LN93,LN92}, in which the transformation laws of the scale variables $\ln \rho$ take the form of a Lorentz group, so  that the fractal dimension becomes itself a variable. In this framework, the differential elements $dX$ can no longer tend to zero since they are limited at small scales by a minimal length scale, invariant under dilations, that we have identified with the Planck length $\lambda_{\Planck}=\sqrt{\hbar G/c^3}$. It has a status similar to that of $c$ in motion special relativity, i.e., of an unreachable and impassable horizon rather than of a cutoff or a barrier: namely, it replaces the zero point since an infinite contraction would be needed to obtain it from another scale. Combined with the role of scale transition played by the Compton length (in rest frame), this interpretation of the Planck scale leads to introduce a set of fundamental constants $\Csc=\ln(\lambda_c/\lambda_{\Planck})=\ln(m_{\Planck}/m)$ which are characteristic of elementary particles of mass $m$ and Compton length $\lambda_c=\hbar/m c$. These constants  play an essential role in structuring the geometry of  the geodesics families of the fractal space-time \cite{LN96} (to which we identify the particles, see next Sec.~\ref{sec:geodesics}), in particular when accounting for the coupling between scale and motion that leads to the emergence of gauge fields, which is the main subject of the present paper.
%end of added paragraph

% 2.3***********************************************
\subsection{Geodesics of a fractal space-time}
\label{sec:geodesics}
%************************************************

%2.3.1**********************
\subsubsection{Infinity of geodesics}
%********************************
The next step in such a geometric approach consists of writing the geodesics equation. We make the conjecture that the description of quantum particles can be reduced to that of these geodesics.  Then their internal properties are the 
geometrical properties of the geodesics bundle corresponding to their state, according to the various conservative quantities (prime integrals) that define them.

Any measurement performed on the ``particle'' is interpreted as a selection 
of the geodesics bundle linked to the interaction with the measurement apparatus (that depends on its resolution) and/or to the information known about it (for example, the which-way-information in a two-slit experiment \cite{LN96}).

Generalizing to space-times the definition of fractal functions, we have 
defined a fractal space-time as the equivalence class of a family of Riemanian 
manifolds, explicitly depending on the scale variables. In such 
space-times, the geodesics equations are also scale dependent and the number 
of geodesics that relate any two events (or starts from any event) is infinite. We are therefore led to adopt a generalized 
statistical fluidlike description where the deterministic velocity $V^{\mu}(s)$ is 
replaced by a scale-dependent, fractal velocity field $V^{\mu}[X^{\mu}(s,ds),s,ds]$.

% 2.3.2**********************
\subsubsection{Discrete symmetry breaking}
%********************************

Another consequence of nondifferentiability is the breaking of the invariance by reflexion of the differential element $ds$. Indeed, for fractal functions $f(s,ds)$, two generalized derivatives are defined instead of one,
\begin{equation}
f'_+(s,ds) = \frac{f(s+ds,ds)-f(s,ds)}{ds}, \;\; 
f'_-(s,ds) = \frac{f(s,ds)-f(s-ds,ds)}{ds},
\end{equation}
that are transformed one into the other by the reflexion $ds \leftrightarrow -ds$. Applied to the space-time coordinates, these two derivatives give two divergent
velocity fields, $V_{+}^{\mu}[x(s,ds),s,ds]$ and $V_{-}^{\mu}[x(s,ds),s,ds]$. Each of them can be in turn decomposed in terms of  classical parts $v_+$ and $v_-$, and of fractal parts $w_+$ and $w_-$.

Then we define two ``classical'' derivatives $d_{+}/ds $ and $d _{-}/ds$,
which, when they are applied to $x^\mu$, yield the ``classical'' velocity fields
\begin{equation}
{\frac{d _{+}}{ds}} x^{\mu}(s)  = v^{\mu} _{+} \qquad
{\frac{d _{- }}{ds}} x^{\mu}(s)  = v^{\mu} _{-}.
\end{equation}
Since there is no reason to privilegize one process rather than the other, we consider both $(+)$ and  $(-) $ processes on the same footing, and we combine them in a unique twin process in terms of which the microscopic reversibility is recovered \cite{LN93}. A simple and natural way to account for this doubling is to use complex numbers and the complex product \cite{CN04}. In the scale relativity framework, this fundamental two-valuedness implied by nondifferentiability can be shown to be the origin of the complex nature of the wave function of 
quantum mechanics.

%2.3.3**********************
\subsubsection{Quantum-covariant derivative}
\label{ccdo}
%********************************
The next step of the scale-relativity program amounts to include in the construction of a complex derivative operator the various effects (described above) of nondifferentiability and fractality \cite{LN93} : 
\begin{equation}
\frac{\dfr}{ds} = {1\over 2} \left( \frac{d_+}{ds} + \frac{d_-}
{ds} \right) - {i\over 2} \left(\frac{d_+}{ds} - \frac{d_-}{ds}\right).
\end{equation}
Such an operator will play the role of a covariant derivative (in Einstein's general meaning given to this word, i.e., as a tool of implementation of the principle of covariance, according to which the fundamental equations of physics should keep their form under transformations of the reference system). It can be used to define a complex four-velocity field,
\begin{equation}
{\cal V}  ^{\mu } =    \frac{\dfr}{ds}  x ^{\mu } = V ^{\mu  }- 
i  U ^{\mu  } =    \frac{v ^{\mu } _{+ }+ v ^{\mu } _{- }}{2}   -  i    
\frac{v ^{\mu } _{+ }- v ^{\mu } _{- }}{2}.
\end{equation}
The total derivative of a fractal function contains finite terms up to 
highest orders. For a constant fractal dimension $D_{F}=2$, a finite contribution 
only proceeds from terms up to second order. Since only stationary functions which do not depend explicitly on $s$ are considered, one can show that the complex covariant derivative operator reads in the relativistic case \cite{LN94,CN04}
\begin{equation}
\label{AAA}
\frac {\dfr}{ds}   =    ({\cal V}  ^{\mu }  +  i  \frac{ \lambda_c}{2} \, 
\partial   ^{\mu } )\, \partial  _{\mu }.
\end{equation}
Finally, using the strong covariance principle (extended to scales), we are led to write a geodesics equation by using this covariant derivative in terms of a freelike equation of motion:
\begin{equation}
\frac {\dfr}{ds}   {\cal V}  ^{\mu } =0.
\end{equation}
At this stage, the wave function $\psi = e^{iS/m \lambda_c}$ is defined
as a mere re-expression of the complex action $S$. By introducing it in the above geodesics equation thanks to its relation to the velocity field $ {\cal V}_{\mu}=i \lambda \,  \partial_{\mu} \ln \psi$, it gives after integration the complex (standard) 
free Klein-Gordon equation, $\lambda_c^2 \,  \partial^{\mu}  \partial_{\mu} \psi + {\psi}=0$
\cite{LN94,LN96}.

%paragraph completed to account for referee's comments
Note that we consider in this paper only the full relativistic case in which both space and time are fractal (which corresponds to energies larger than $m c^2$, i.e., to scales smaller than the Compton scale). However it is worth briefly recalling that  nonrelativistic quantum mechanics (which usually applies at intermediate scales) is recovered in our framework in terms of a three-dimensional fractal space, with no fractal time.  In this case a generalized Schr\"odinger equation for a complex wave function is derived \cite{LN93,LN96,CN04,LN97b,DRN03,JC03}. The reason for such an asymmetry between space and time in the scale-relativity description (and in quantum mechanics) is to be found in the quantum-classical transition, identified with the fractal-nonfractal transition (see Sec.~\ref{sec:metric}). Indeed, for a free particle it is given by the Einstein-de Broglie scale $\lambda_{\mu}=\hbar/p^{\mu}$, whose time scale $\tau=\hbar/E$ is always smaller than its corresponding space-scales $\lambda=\hbar/p$, because of the relation $E^2=p^2+m^2$ (and therefore finally because of the existence of mass). This implies a first transition from standard space-time to fractal space, then at smaller scales a second transition to fractal space-time \cite{LN93}. These three regimes manisfest themselves successively as classical, quantum non-relativistic then quantum relativistic mechanics. 
%end of completed paragraph

%2.4**********************

\subsection{The Dirac equation as a geodesics equation in a fractal space-time}
\label{sec:dirac}
%********************************

As recalled hereabove, the Klein-Gordon equation is obtained as a result of the $ds \leftrightarrow 
-ds$ symmetry breaking. The consideration of a more general case where we add 
the breaking of the symmetries $dx^{\mu}\leftrightarrow -dx^{\mu}$ and 
$x^{\mu}\leftrightarrow -x^{\mu}$ leads to the appearance of  bispinors which are solutions of the Dirac equation \cite{CN04}.

Following the method described in Sec.~\ref{ccdo}, these additional discrete symmetry breakings lead to two new doublings of the velocity field and of the classical derivative. The four-velocity field has now eight components, which are used to construct a 
biquaternionic (complex-quaternionic) velocity field. Then a  biquaternionic covariant derivative operator may be built, which keeps once again the same form (Eq.~\ref{AAA}) as in the complex case \cite{CN04}, even though the velocity field is now a biquaternion instead of a complex number.

The biquaternionic geodesics equation reads
\begin{equation}
\frac{\dfr}{ds} \,{\cal V}_{\alpha}= ({\cal V}  ^{\mu }  +  i  \frac{ \lambda_c}{2} \, 
\partial   ^{\mu } )\; \partial  _{\mu } {\cal V}_{\alpha}=0.
\label{eq.bqge}
\end{equation}
 A biquaternionic action is defined according to
\begin{equation}
\delta {\cal S}= \partial_{\mu}{\cal S} \; \delta x^{\mu}=- mc \; 
{\cal V}_{\mu} \; \delta x^{\mu}.
\end{equation}
The biquaternionic four-momentum can therefore be written
${\cal P}_{\mu}=mc \; {\cal V}_{\mu}= -\partial_{\mu}{\cal S}$.
Then we introduce a biquaternionic wave function, which 
is once again a mere re-expression of the action, as
\begin{equation}
\psi^{-1} \partial_{\mu} \psi = {i\over {c S_0}} \partial_{\mu} 
{\cal S},
\end{equation}
which yields for the 
biquaternionic four-velocity the expression
\begin{equation}
{\cal V}_{\mu}=i \lambda_c \psi^{-1} \partial_{\mu} \psi.
\label{eq.bqfv}
\end{equation}
This relation is destined to play an essential role in the subsequent construction of the non-Abelian gauge theory. Indeed, its specific form ($ \psi^{-1} \partial_{\mu} \psi$), which is linked to the noncommutativity of biquaternions, will allow a proper generalization to multiplets which permits in its turn a geometric construction of the non-Abelian charges in accordance with the standard Yang-Mills theory (Sec.~\ref{s.nagf}).

Then we replace in Eq.~(\ref{eq.bqge}) the 
velocity field ${\cal V}_{\alpha}$ by its expression (Eq.~\ref{eq.bqfv}). We obtain the motion equation as a third order differential equation, which becomes after some calculations $\partial_{\mu}[(\partial^{\nu}\partial_{\nu} \psi) \psi^{-1}] = 0$ and may therefore be integrated. This yields the 
Klein-Gordon equation for a free particle, $\lambda_c^2 \,  \partial^{\mu}  \partial_{\mu} \psi + {\psi}=0$, but now generalized to complex 
quaternions \cite{CN03,CN04}. 

Long-known properties of the quaternionic formalism (see, e.g., \cite{CL29,AC37}) can finally be used  to readily obtain the Dirac equation for a free particle, namely,
\begin{equation}
{1\over c}{\partial \psi \over {\partial t}} = - \alpha^k{\partial \psi \over 
{\partial x^k}} - i{mc\over \hbar}\beta \psi ,
\label{eq.102}
\end{equation}
as a mere square root of the Klein-Gordon operator \cite{CN03,CN04}, which was itself derived from the geodesics equation~(\ref{eq.bqge}).
Then the isomorphism which can be established between the quaternionic and  
spinorial algebrae \cite{PR64} allows us to identify the wavefunction $\psi$ 
to a Dirac spinor. In a Lagrangian formalism, the Dirac equation proceeds from the Lagrangian 
density, 
\begin{equation}
{\cal L}=  \bar{ \psi} \; (i \gamma ^{\mu} \partial_{\mu} -m )\;  \psi,
\label{3010}
\end{equation}
which therefore becomes a direct consequence of the scale relativity 
principles \cite{CN04}. It is also easy to derive the Pauli equation, since it is known that it 
can be obtained as a non-(motion)-relativistic approximation of the Dirac 
equation, while, in this approximation, Dirac bispinors become Pauli spinors. 

%paragraph added to account for the referee's comment
Let us conclude this section by a final remark: one of the consequences of this theory is that it provides a physical picture of the nature of spin. In the scalerelativistic framework, the complex nature of the wave function and the existence of spin have both a common origin, namely, the fundamental twovaluedness of the derivative (in its generalized definition) coming from nondifferentiability. These two successive doublings are naturally accounted for in terms of algebra doublings (see Appendix of Ref.~\cite{CN04}), i.e., of a description tool that jumps from real numbers ${\rm I\! R}$ to complex numbers ${\rm I\! \!\!C}={\rm I\! R}^2$, then to quaternions ${\rm I\! H}={\rm I\! \!\!C}^2$. However, while the origin of the complex nature of the wave function is linked to the total derivative (and therefore to proper time) through the doubling ${d}/{ds}\rightarrow ({d_+}/{ds},{d_-}/{ds})$, the origin of spin is linked to the partial derivative with respect to the coordinates through the doubling ${\d}/{\d x^{\mu}}\rightarrow ({\d_+}/{\d x^{\mu}},{\d_-}/{\d x^{\mu}})$, which finally leads to the twovaluedness of the wave function itself $\psi \rightarrow (\psi_1,\psi_2)$, characterizing a (Pauli) spinor. 

Such a  physical effect has naturally a consequence on the angular momentum $(x^{\mu}{\d}/{\d x^{\nu}}-x^{\nu}{\d}/{\d x^{\mu}})$, leading to the two directions in which spin can become locked. Moreover, numerical simulations of the fractal geodesical  curves (work in preparation) which are solutions of Eq.~(\ref{eq.bqge}) allow us to obtain a more specific picture of the spin as an internal angular momentum of these geodesics. Indeed, these solutions are characterized by spiral structures at all scales, in agreement with Ord's reformulation of the Feynman relativistic chessboard model in terms of spiral paths \cite{G092}. They also support our early models of emergence of a spinlike internal angular momentum in fractal spiral curves of fractal dimension 2 \cite{LN93,LN89}. We here recall briefly the argument: the angular momentum $L_z=m r^2\dot{ \varphi}$ should classically vanish for $r\rightarrow 0$. But in the fractal spiral model, $\dot{ \varphi}\rightarrow \infty$ when  $r\rightarrow 0$ in such a way that $ r^2\dot{ \varphi}$ remains finite when $D_F=2$ (while it is vanishing for $D_F<2$ and divergent for $D_F>2$). This result solves the problem of the apparent impossibility to define a spin in a classical way both for an extended object and for a pointlike object, and provides another proof of the critical character of the value $D_F=2$ for the fractal dimension of quantum particle paths (which can be derived from the Heisenberg relations).
%end of added paragraph

%3*******************************
\section{Scale-relativistic theory of electromagnetism: summary}
\label{s.elec}
%********************************

%3.1****************************************

\subsection{ Electromagnetic field and electric charges}

%******************************************

Let us now briefly recall the results previously obtained 
in the case of a U(1) field \cite{LN94,LN96,LN03}. 

We consider here a special situation in which the set of the scale variables 
comes down to only one element, $\varrho=\varepsilon/\lambda$. This amounts to limit ourselves to the study of global 
scale transformations (contractions/dilations) in ``scale-space''. 

Because, according to the principle of scale relativity, this ``scale-space'' is fundamentaly non-absolute, the
scale of a structure (internal to the fractal geodesics which are identified with a ``particle") is expected to change during a displacement in space-time. In other words, we now consider scale variables which become explicit functions of the coordinates, i.e., $\varrho=\varrho(x,y,z,t)$.

This is analogous to the situation encountered in general relativity (GR) for 
a curved space-time: namely, in a parallel displacement, a vector $V^{\mu }$
is subjected to an increase $\delta V^{\mu }=-\Gamma _{\nu \rho }^{\mu
}V^{\nu }dx^{\rho }$ (where the $\Gamma _{\nu \rho }^{\mu }$ are the Christoffel symbols, i.e., the gravitational field), due to the geometric effects of curvature. Then, if one
substracts this geometric increase from its total variation $dV^{\mu }$, one
recovers the inertial part of the variation (see, e.g., \cite{LL2}).
This allows to define the GR covariant derivative $D$ as $DV^{\mu }=dV^{\mu }-\delta V^{\mu }=dV^{\mu }+\Gamma _{\nu \rho }^{\mu
}V^{\nu }dx^{\rho }$.

The same kind of behavior is true in the scale relativity framework, but
with an essential difference: while the effects of curvature affect
vectors, tensors, etc., but not scalars, the effects of fractality begin
already at the level of scalars, among which the ``invariant'' of length $
ds^{2}$ itself.

Therefore, we expect in a displacement the appearance of a
resolution change due to the fractal geometry, that
reads
\begin{equation}
\delta \varepsilon =-\frac{1}{q}\;A_{\mu }\;\varepsilon \;dx^{\mu },
\label{0002}
\end{equation}
i.e., in terms of the scale ratio, 
\begin{equation}
 \label{eq3}
\delta \ln \varrho =\frac{1}{q}\;A_{\mu }\;dx^{\mu }. 
\end{equation}

The introduction of the $(1/q)$ term in this definition is an important
point for the electromagnetic case and also for its non-Abelian
generalizations. Indeed, as we shall see
in what follows, the ``field" $A_{\mu}$ will be identified with an
electromagnetic potential. Since $\ln \varrho$ is dimensionless, we are led
to divide the potential term by the ``active'' electric charge $q$, leaving
a charge-independent purely geometric contribution.

This leads to the appearance of a dilation field, according to the
construction of a scale-covariant derivative,
\begin{equation}
D\chi =d\chi -\delta \chi =d\chi -A_{\mu }dx^{\mu },
\end{equation}
where we have set $\chi =q\ln \varrho$. We finally obtain the partial derivative as the sum of the inertial and of the geometric terms as 
\begin{equation}
\partial _{\mu }\chi =D_{\mu }\chi +A_{\mu }.  \label{eq2}
\end{equation}

Let us now consider the action $S$ for, e.g., an electron. In the framework of a
space-time theory based on a relativity principle, which is here the case,
its variation should be given directly by the space-time invariant $ds$, i.e., 
$\delta \int dS=0$ becomes identical with a geodesics (Fermat) principle $
\delta \int ds=0$. But now the fractality of the geodesical curves to which
the electron wave field is identified means that their proper length becomes
a function of the scale variable, so that $S=S(\chi )$.

Therefore the differential of the action reads
\begin{equation}
dS=\frac{\partial S}{\partial \chi }\,d\chi =\frac{\partial S}{\partial \chi
}\,(D\chi +A_{\mu }dx^{\mu }),  \label{2222}
\end{equation}
so that we obtain
\begin{equation}
\partial _{\mu }S=D_{\mu }S+\frac{\partial S}{\partial \chi }\;A_{\mu }.
\end{equation}
This result provides us with a definition for the ``passive'' charge (on
which the electromagnetic field acts) as \cite{LN94,LN96}
\begin{equation}
\frac{e}{c}=-\frac{\partial S}{\partial \chi }.  \label{5893}
\end{equation}
This is a second important point worth to be emphasized, since it will play
an important role for the generalization to non-Abelian gauge 
theories. In the standard theory, the charge is set from experiment, then it
is shown to be related to gauge transformations, while the gauge functions
are considered to be arbitrary and devoid of physical meaning. In the scale
relativity approach, the charges are built from the symmetries of the ``scale
space''. One indeed recognizes in Eq.~(\ref{5893}) the standard expression that
relates a conservative quantity to the symmetry of a fundamental variable
(here, the relative resolution), according to Noether's theorem. 

Note that, at this level of the construction of the theory, the charge is defined as a large scale prime integral (conservative quantity). But, once this result is obtained, a second step consists of studying in detail the internal structures of the fractal geodesics (that are identified with the charged particle) \cite{LN96,LN93}. These internal structures can afterwards be interpreted (at time scales smaller than $\tau=\hbar/E$), in terms of virtual particle-antiparticle pairs, then of radiative corrections and of the scale variation of the charge toward small scales, as described by the renormalization group equations.

It is also remarkable that, in such a relativistic foundation of electromagnetism, we are led to introduce in a separate way an active and a passive charge.  This is also analogous to  the introduction in GR of an active gravitational mass and of a passive mass which are equal according to the GR strong principle of equivalence. As a consequence, a scale-relativistic principle of equivalence of these charges can be set (in order to account for the action-reaction principle in Coulomb's law). Under this principle, $e=q$ and Eq.~(\ref{5893}) becomes $e^2/c=-\partial S/\partial \ln \varrho$. 

We have therefore established from first principles the form of the
action in the classical electromagnetic theory, in particular the form of
the particle-field coupling term (which was postulated in the standard theory), as (see, e.g., \cite{LL2})
\begin{equation}
dS=-mc\,ds-\frac{e}{c}\,A_{\mu }\,dx^{\mu }.
\end{equation}
But this form has also a new geometrical interpretation. It means that, in this framework, an
increase of the length can come from two contributions: the first is the
usual variation due to the motion of the particle, while the second new
contribution, which is of geometric nature, is a length dilation of the internal fractal
structures.

We are now able to write a geodesics equation minimizing the
length invariant (i.e., the proper time), which coincides with the
least-action principle $\delta \int dS=0$ (see \cite{LL2}). The
variation of the above action yields the Lorentz equation of
electrodynamics,
\begin{equation}
mc\,\frac{du_{\alpha }}{ds}=\frac{e}{c}\,F_{\alpha \mu }\,u^{\mu },
\end{equation}
where $F_{\alpha \mu
}=\partial _{\alpha }A_{\mu }-\partial _{\mu }A_{\alpha }$ is the electromagnetic tensor field. 
We also recover the standard form for the differential of the action as
a function of the coordinates, namely,
\begin{equation}
dS=-(mc\,u_{\mu }+\frac{e}{c}\,A_{\mu })\,dx^{\mu }.  \label{9237}
\end{equation}

%*****************************

\subsection{Quantum electrodynamics}

%******************************

Let us proceed with a brief account of the generalization of this approach
to quantum electrodynamics. As recalled in Sec.~\ref{s.de}, in the scale relativistic approach to the quantum theory
\cite{LN96,LN93}, the four-velocity $\mathcal{V}^{\mu }$ that
describes a scalar particle is complex, so that its action is also a complex
number and it now writes $S=S(x^{\mu },\mathcal{V}^{\mu },\chi )$. The wave function 
is defined from this action as $\psi =\exp \left( i{S}/{\hbar }\right)$.

Therefore Eq.~(\ref{9237}) now takes the form
\begin{equation}
dS=-mc\mathcal{V}_{\mu }\;dx^{\mu }-\frac{e}{c}A_{\mu }dx^{\mu }.
\label{2763}
\end{equation}

The new relation between the wave function and the velocity reads
\begin{equation}
mc\mathcal{V}_{\mu }=i\hbar \;D_{\mu }\ln \psi =i\hbar \partial _{\mu }\ln
\psi -\frac{e}{c}A_{\mu },
\end{equation}
so that we recover the standard QED-covariant derivative as being
nothing but the scale-covariant derivative previously introduced, but now
acting on the wave function,
\begin{equation}
D_{\mu }=\partial _{\mu }+i\,\frac{e}{\hbar c}\,A_{\mu }.
\end{equation}

%3.3********************

\subsection{Gauge invariance}

%**********************

Let us now consider a second internal structure of the fractal geodesics, that
lies at a relative scale $\varepsilon ^{\prime }=\rho ^{\prime }\lambda $.
Equation (\ref{eq3}) becomes
\begin{equation}
\delta \ln \rho ^{\prime }=\frac{1}{q}A_{\mu }^{\prime } \,dx^{\mu }.
\end{equation}
Let $\varphi $ be the ratio between the scales $\varepsilon ^{\prime }$ and $%
\varepsilon $. In the framework of Galilean scale relativity (where the product of two successive dilations $\rho$ and $\rho'$ is $ \rho^{\prime\prime}=\rho \times \rho '$), this ratio is simply
$\varphi ={\rho ^{\prime }}/{\rho }$. One therefore finds 
\begin{equation}
A_{\mu }^{\prime }=A_{\mu }+q\,\partial _{\mu }\ln \varphi ,
\end{equation}
which is the standard gauge invariance relation for the potential. But here a gauge transformation, instead of being arbitrary, is identified with a scale
transformation of the resolution variable in scale space. Under such a transformation, the wave function of the particle becomes
\begin{equation}
\label{2579}
\psi ^{\prime }=\psi \;\exp \left( -i\frac{eq}{\hbar c}\ln \varphi \right) .
\end{equation}
As a consequence, the Lagrangian given by Eq.~(\ref{2763}), that includes the particle and  field-particle coupling terms,
remains globally invariant under a gauge transformation.

When $q=e$ (the electron charge), we have $e^{2}=4\pi \alpha \hbar c$, where
$\alpha $ is the ``fine structure constant'', i.e., the electromagnetic
coupling constant. The previous expression becomes in this case 
$\psi ^{\prime }=\psi \;\exp (-i4\pi \alpha \ln \varphi )$.
 %added in revised version
 In the framework of the special scale-relativity theory \cite{LN92} in which possible scale ratios become limited [$\ln \varphi < \ln (m_{\Planck}/m_e)$] because of the identification of the Planck length scale with a lowest (invariant under dilations) scale,
 %end of added text
this expression has been used to suggest the existence of a relation between the mass and the charge of the electron \cite{LN94,LN96,LN03}.

Let us conclude this review part by stressing that the scale-relativity theory of electromagnetism shares some features with the Weyl-Dirac theory \cite{weyl,dirac}, but that it has new and essential differences.
Namely, the Weyl theory considers scale transformations of the line element,
$ds\rightarrow ds^{\prime }=\rho\, ds$, but without specifying any fundamental
cause for this dilation. The variation of $ds$ should therefore
exist at all scales, in contradiction with the observed invariance of the
Compton length of the electron (i.e., of its mass).

In the scale relativity proposal, the change of the line element comes from
the fractal geometry of space-time, and it is therefore a consequence of the
dilation of the scale variables (``resolutions"). Moreover, the explicit effects of the dependence on
resolutions is observable only below the transition between scale dependence and 
scale independence, which is identified with the Compton scale of the particle 
in its rest frame. This ensures the invariance, in this theory, of the observed 
electron mass.

%4**********************

\section{Non-Abelian gauge fields}
\label{s.nagf}

%********************************

%4.1****************************

\subsection{Scale-relativistic description}

%******************************

%4.1.1****************************

\subsubsection{Introduction}

%******************************

We now generalize the electromagnetic description to a geometric foundation of non-Abelian gauge theories, based upon the scale relativity first principles. 
We consider that the internal fractal structures of the ``particle'' (i.e., of 
the family of geodesics of a nondifferentiable space-time) are now described 
in terms of several scale variables $\eta _{\alpha \beta \ldots }(x,y,z,t)$, that generalize the single 
resolution variable $\varepsilon $. We write them for simplicity in units of $\lambda $, and
we assume that the various indices can be gathered into one common index: we
therefore write the scale variables under the simplified form $\eta _{\alpha
}$ ($\alpha =0$ to $N$).

In the simplest case, $\eta _{\alpha }=\varepsilon _{\alpha }$, where 
$\varepsilon _{\alpha }$ correspond to the resolutions of the space-time 
coordinates $X_{\alpha }$ ($\alpha =1$ to $4$). However, other situations can 
be considered, since their true nature is tensorial rather than vectorial, and since, in analogy with GR, general transformations can be applied to these variables, for example, the transformation $\varepsilon _{\alpha } \rightarrow \ln\varepsilon _{\alpha }$ may be particularly relevant for such scale variables. In this paper, we shall not be more specific about the choice of the scale variables, in order to keep generality. Moreover, our aim here is mainly to relate in a general way 
the scale-relativistic tools to the standard description of current gauge 
theories, so that we shall present only a general description of the scale 
transformations obtained, leaving to future works a more specific establishment of the final gauge group. However, even at this preliminary stage of the analysis, we can show that in any case it contains at least an SU(2) subgroup, e.g.,  the three-dimensional rotations in scale-space which
can be identified with the isospin transformation group (see Sec.~\ref{ss.rss} below). 

%4.1.2****************************

\subsubsection{General scale transformations}

%******************************

Let us consider infinitesimal scale transformations. The transformation 
law on the $\eta _{\alpha }$ can be written in a linear way as
\begin{equation}
\eta _{\alpha }^{\prime }=\eta _{\alpha }+\delta \eta _{\alpha }=(\delta
_{\alpha \beta }+\delta \theta _{{\alpha }{\beta }})\,\eta ^{\beta },
\end{equation}
where $\delta _{\alpha \beta }$ is the Kronecker symbol, or equivalently,
\begin{equation}
\delta \eta _{\alpha }=\delta \theta _{{\alpha }{\beta }}\;\eta ^{\beta }.
\end{equation}
Let us now assume that the $\eta _{\alpha }$'s are functions of the standard
space-time coordinates. This leads us to generalize the scale-covariant
derivative previously defined in the electromagnetic case as follows: the
total variation of the resolution variables becomes the sum of the inertial
one, described by the covariant derivative, and of the new geometric
contribution, namely,
\begin{equation}
d\eta _{\alpha }=D\eta _{\alpha}-\eta ^{\beta}\delta \theta _{{\alpha }{%
\beta }}=D\eta _{\alpha}-\eta ^{\beta}W_{{\alpha }{\beta }}^{\mu
}\;dx_{\mu }.
 \label{0027}
\end{equation}
Note that, here, this covariant derivative is similar to that of GR, i.e., it amounts to subtract the new geometric part in order to keep only the inertial part (for which the motion equation will therefore take a geodesical, freelike form). This is different from the case of the quantum-covariant derivative (Eq.~\ref{AAA}), which includes the effects of nondifferentiability by adding new terms in the total derivative.

Recall that in the Abelian case, which corresponds to a unique global
dilation, this expression can be simplified since $d\eta /\eta =d\ln \eta
=d\chi $. We want also to note here that we have chosen  to 
write the new geometric contribution $-\eta ^{\beta}\delta \theta _{{\alpha }
{\beta }}$, i.e., with a minus
sign, in order to recover the covariant 
derivative of gauge theories in its standard form (this is actually an inessential sign ambiguity).

In this new situation we are led to introduce ``gauge field potentials'' $W_{{\alpha }{\beta }}^{\mu }$ that enter naturally in the 
geometrical frame of Eq.~(\ref{0027}). These potentials  are linked to the scale 
transformations as follows:
\begin{equation}
\delta \theta _{{\alpha }{\beta }}=W_{{\alpha }{\beta }}^{\mu }\;dx_{\mu }.
\end{equation}
One should remain cautious about this expression and keep in mind that these
potentials find their origin in a covariant derivative process and are therefore
not gradients (this is expressed by the use of a difference sign $\delta
\theta _{{\alpha }{\beta }}$ instead of $d\theta _{{\alpha }{\beta }}$). 
They formalize the coupling between displacements in space-time and 
transformations of the scale variables and play in Eq.~(\ref{0027}) 
a role analogous to the one played in general relativity by the Christoffel 
symbols. It is also important to notice that the $W_{{\alpha }{\beta }}^{\mu }$ introduced at this level of the analysis do not
include charges. They are functions of the space and time coordinates only.
This is a necessary choice because our method generates, as we shall see, 
not only the fields but also the charges from, respectively, the
scale transformations and the scale symmetries of the dynamical fractal
space-time.

%*************************
\subsubsection{Multiplets}

After having written the transformation law of the basic variables
(the $\eta _{\alpha }$'s), we are now led to describe how various physical
quantities transform under these $\eta _{\alpha }$ transformations. These new
transformation laws are expected to depend on the nature of the objects
to transform (e.g., vectors, tensors, spinors, etc.), which implies to
jump to group representations.

In the case where the particle is a spin-1/2 fermion, it has been recalled in 
Eq.~(\ref{eq.bqfv}) that the relation between the velocity and the spinor 
fields reads
\begin{equation}
\mathcal{V}_{\mu }=i\lambda \;\psi ^{-1}\partial _{\mu }\psi ,
\end{equation}
where $\mathcal{V}_{\mu }$ and $\psi $ are complex quaternions and the 
constant $\lambda =\hbar /mc$ is the Compton length of the particle.

However, bispinors are not a general enough description for fermions
subjected to a general gauge field. Indeed, we consider here a generalized
group of transformations which therefore involves generalized charges. As a
consequence of these new charges (whose existence will be fully justified below and their form specified), the very nature of the fermions is
expected to become more complicated. Experiments have indeed shown that new
degrees of freedom must be added in order to represent the weak isospin,
hypercharge and color. In order to account in a general way for this more
complicated description, we shall simply introduce multiplets $\psi _{k}$,
where each component is a Dirac bispinor. Therefore, as already remarked in previous
presentations \cite{LN96}, when the scale variables become multiplets, the
same is true of the charges. As we shall see in what follows, in the present
approach it is at the level of the construction of the charges that the
set generators enter.

In this case the multivalued velocity becomes a biquaternionic matrix,
\begin{equation}
\mathcal{V}_{jk}^{\mu }=i\lambda \;\psi _{j}^{-1}\partial ^{\mu }\psi _{k}.
\end{equation}
The biquaternionic (therefore noncommutative) nature of the wave function (which is equivalent to Dirac bispinors) plays here an essential role, as previously announced. Indeed, it leads to write the velocity field as $\psi^{-1}\partial ^{\mu }\psi$ instead of $\partial ^{\mu }\ln \psi$ in the complex case, so that its generalization to multiplets involves two indices instead of one. As we shall see in what follows, the general structure of Yang-Mills theories and the correct construction of non-Abelian charges will be obtained thanks to this result.

Therefore the action becomes also a tensorial two-index quantity,
\begin{equation}
dS_{jk}=dS_{jk}(x^{\mu },\mathcal{V}_{jk}^{\mu },\eta _{\alpha }).
\end{equation}
In the absence of a field, it is linked to the generalized velocity (and
therefore to the spinor multiplet) by the relation
\begin{equation}
\partial ^{\mu }S_{jk}=-mc\;\mathcal{V}_{jk}^{\mu }=-i\hbar \;\psi _{j}^{-1}%
\partial ^{\mu }\psi _{k}.  \label{5307}
\end{equation}

Now, in the presence of a field [i.e., when the second-order 
effects of the fractal geometry appearing in the right hand side of 
Eq.~(\ref{0027}) are included], using the
complete expression for $\partial ^{\mu }\eta _{\alpha }$,
\begin{equation}
\partial ^{\mu }\eta _{\alpha }=D^{\mu }\eta _{\alpha }-W_{\alpha \beta
}^{\mu }\;\eta ^{\beta },
\end{equation}
we are led to write a relation that generalizes Eq.~(\ref{2222}) to the
non-Abelian case,
\begin{equation}
\partial ^{\mu }S_{jk}=\frac{\partial S_{jk}}{\partial \eta _{\alpha }}\;%
\partial ^{\mu }\eta _{\alpha }=\frac{\partial S_{jk}}{\partial \eta
_{\alpha }}\,(D^{\mu }\eta _{\alpha }-W_{\alpha \beta }^{\mu }\;\eta ^{\beta
}).
\end{equation}
Thus we obtain
\begin{equation}
\partial ^{\mu }S_{jk}=D^{\mu }S_{jk}-\eta ^{\beta }\;\frac{\partial S_{jk}}{%
\partial \eta _{\alpha }}\;W_{\alpha \beta }^{\mu }.  \label{1256}
\end{equation}

We are finally led to define a general group of scale transformations whose
generators are
\begin{equation}
T^{\alpha \beta }=\eta ^{\beta }\partial ^{\alpha }
\end{equation}
(where we use the compact notation $\partial ^{\alpha }=\partial /\partial
\eta _{\alpha }$), yielding the generalized charges,
\begin{equation}
\frac{\tilde{g}}{c}\;t_{jk}^{{\alpha }{\beta }}=\eta ^{\beta }\;\frac{%
\partial S_{jk}}{\partial \eta _{\alpha }}.
\end{equation}
This group is submitted to a unitarity condition, since, when it is applied
to the wave functions, $\psi \psi ^{\dagger }$ must be conserved.

%**************************************

\subsubsection{Rotations in ``scale-space''}
\label{ss.rss}

%**************************************

In order to enlight the meaning of the new definition we have obtained for
the charges, we consider in the present section a subsample of the
possible scale transformations on intrinsic scale variables: namely,
those which are built from the
antisymmetric part of the gauge set (that can therefore be identified as rotations in the scale space).  In this case the infinitesimal
transformation is such that
\begin{equation}
\delta \theta _{{\alpha }{\beta }}=-\delta \theta _{{\beta }{\alpha }%
}\Rightarrow W_{{\alpha }{\beta }}^{\mu }=-W_{{\beta }{\alpha }}^{\mu }.
\end{equation}
Therefore, reversing the indices in Eq. (\ref{1256}), we may write
\begin{equation}
\partial _{\mu }S_{jk}=D_{\mu}S_{jk}-\eta ^{\alpha }\;\frac{\partial S_{jk}%
}{\partial \eta _{\beta }}\;W_{{\beta }{\alpha }}^{\mu }.  \label{4573}
\end{equation}
Taking the half-sum of Eqs. (\ref{1256}) and (\ref{4573}) we finally obtain
\begin{equation}
\partial _{\mu }S_{jk}=D_{\mu }S_{jk}-\frac{1}{2}\left( \eta ^{\beta }\;%
\frac{\partial S_{jk}}{\partial \eta _{\alpha }}-\eta ^{\alpha }\;\frac{%
\partial S_{jk}}{\partial \eta _{\beta }}\right) \;W_{{\alpha }{\beta }%
}^{\mu }.
\end{equation}

This leads to define the new charges,
\begin{equation}
\frac{\tilde{g}}{c} \; t^{{\alpha}{\beta}}_{jk}=\frac{\partial S_{jk}}{%
\partial \theta_{{\alpha}{\beta}}} = \frac{1}{2}\left(\eta^{\beta} \; \frac{%
\partial S_{jk}}{\partial \eta_{\alpha}} - \eta^{\alpha}\; \frac{\partial
S_{jk}}{\partial \eta_{\beta}}\right).
\end{equation}
We recognize here a definition similar to that of the angular momentum, i.e., of the conservative quantity that finds its
origin in the isotropy of space; but the space under consideration is here
the ``scale-space'', i.e., the space of the scale variables that must be added for a proper description of a fractal geometry of space-time. Therefore the charges of the gauge fields are identified,
in this interpretation, with ``scale-angular momenta''.

The subgroup of transformations corresponding to these generalized charges
is, in three dimensions, a SO(3) group related to a SU(2) group by the
homomorphism which associates to two distinct 2$\times $2 unitary matrices
of opposite sign the same rotation. We are therefore naturally led to define
a ``scale-spin'', which we propose to identify to the simplest non-Abelian
charge in the current standard model: the weak isospin.

Coupling this SU(2) representation of the rotations in a three dimensional 
sub-``scale-space'' to the
U(1) representation of the global scale dilations (that describes, e.g.,  the
electromagnetism process) analyzed in Sec.~\ref{s.elec}, we are therefore 
able to give a physical geometric meaning to the transformation group corresponding to 
the U(1)$\times$SU(2) representation of the standard electroweak theory \cite{SW67,AS68}.

It is worth stressing here that the group of three-dimensional rotations in ``scale space'' 
is only a subgroup of an at least four dimensional rotation group (one scale 
variable for each space-time coordinate), and therefore at least SO(4), and, 
more precisely, its universal covering group SU(2) $\times$ SU(2).

%**************************************

\subsection{Yang-Mills theory with the scale relativity tools}
\label{s.ymt}

%**************************************

\subsubsection{Simplified notation}

For the subsequent developments, we shall simplify the notations and use only
one index $a=(\alpha ,\beta )$ for the scale transformations: this index runs
on the gauge group parameters, now written $\theta _{a}$. For example, in
three dimensions, this means that we replace the three rotations $\theta
_{23},\theta _{31},\theta _{12}$, respectively, by $\theta _{1},\theta
_{2},\theta _{3}$. We obtain the following more compact form for the
complete action:
\begin{equation}
dS_{jk}=\left( D_{\mu }S_{jk}-\frac{\tilde{g}}{c}\;t_{jk}^{a}\;W_{a\mu
}\right) dx^{\mu },
\end{equation}
and therefore
\begin{equation}
\label{BBB}
D^{\mu }S_{jk}=-i\hbar \;\psi _{j}^{-1}D^{\mu }\psi _{k}=-i\hbar \;\psi
_{j}^{-1}\partial ^{\mu }\psi _{k}+\frac{\tilde{g}}{c}\;t_{jk}^{a}\;W_{a}^{%
\mu }.
\end{equation}

%**************************************

\subsubsection{Scale relativistic tools for Yang-Mills theory}

%**************************************

The previous equations have used new concepts that are specific of the scale
relativity approach, namely (i) the scale variables $\eta_{\alpha}$, (ii) the
biquaternionic velocity matrix $\mathcal{V}_{jk} ^{\mu}$, and (iii) its associated
action $S_{jk}$. The standard concepts of quantum field theories,
namely the fermionic field $\psi$, the bosonic field $W_a^{\mu}$, the charges
$g$, the gauge group generators $t^a_{jk}$ and the gauge-covariant
derivative $D_{\mu}$ are here all of them derived from these new tools.

Let us show that we are thus able to recover the basic relations of standard
non-Abelian gauge theories (see e.g. \cite{IA82}). From Eq.~(\ref{BBB}), we first obtain 
the standard form for the covariant partial derivative, now acting on the
wave function multiplets,
\begin{equation}
D^{\mu} \psi_k= \partial^{\mu} \psi_k + i \, \frac{\tilde{g}}{\hbar c} \;
t^{ja}_{k} \; W_a^{\mu} \; \psi_j.
\end{equation}
The $\psi_k$'s do not commute together since they are biquaternionic
quantities, but this is the case neither of $t^{ja}_{k}$ nor of 
$W_a^{\mu}$,
so that $\psi_j$ can be set to the right as in the standard way of writing;
from the multiplet point of view (index $j$), we simply exchange the lines
and columns.

Now introducing a dimensionless coupling constant $\alpha _{g}$ and a
dimensionless charge $g$, such that
\begin{equation}
g^{2}=4\pi \alpha _{g}=\frac{\tilde{g}^{2}}{\hbar c},
\end{equation}
and redefining the dimensionality of the gauge field (namely, we replace $W_{a}^{\mu }/\sqrt{\hbar c}$ by $W_{a}^{\mu }$), the covariant derivative
may be more simply written under its standard form,
\begin{equation}
D^{\mu }\psi _{k}=\partial ^{\mu }\psi _{k}+i \, g  \;  t_{k}^{ja}  \;  W_{a}^{\mu} \; \psi _{j}, 
 \label{5555}
\end{equation}
 where all of the three new contributions, $g$, $ t_{k}^{ja}$, and $W_{a}^{\mu}$ have been constructed from the origin by the theory and given a geometric meaning.
 
In the simplified case of a fermion singlet, it reads
\begin{equation} \label{5566}
D^{\mu }=\partial ^{\mu }+i\,g\;t^{a}\;W_{a}^{\mu }.
\end{equation}
Let us now derive the laws of gauge transformation for the fermion field.
Consider a transformation $\theta_a$ of the scale variables. As we shall now
see, the $\theta_a$ can be identified with the standard parameters of a
non-Abelian gauge transformation. Indeed, using the above remark about the
exchange of lines and columns, Eq.~(\ref{5307}) becomes
\begin{equation}
-i \hbar \partial^{\mu} \psi_k= \partial^{\mu} S^j_k \; \psi _j 
\end{equation}
and allows us to recover by a different way Eq.~(\ref{5555}), from which 
we obtain the standard form
for the transformed fermion multiplet in the case of an infinitesimal gauge
transformation $\delta \theta_a$,
\begin{equation}
\psi^{\prime}_k= \left( \delta^j_k - i g \, t^{ja}_k \, \delta \theta_a \right) \psi_j .
\end{equation}

%**************************************

\subsubsection{Yang-Mills theory}

%**************************************

We have now at our disposal all the tools of quantum gauge theories. The
subsequent developments are standard ones in terms of these tools. Namely,
one introduces the commutator of the matrices $t_a$ (which have {\it a priori} no
reason to commute), under the form
\begin{equation}
t_a t_b - t_b t_a= i f_{ab}^c \; t_c .
  \label{4011}
\end{equation}
Therefore the $t_a$'s are identified with the generators of the gauge group
and the $i \, f_{ab}^c$'s with the structure constants of its associated Lie
algebra. The noncommutativity of the generators and the requirement of the 
full Lagrangian invariance under the scale transformations finally imply 
the appearance of an additional term in the gauge transformation law of the 
boson fields. We obtain this additional term by the standard method 
recalled below.

We replace into the Lagrangian of the fermionic field given by 
Eq.~(\ref{3010}) the partial derivative $\partial_{\mu}$ by its covariant 
counterpart $D_{\mu}$ of Eq.~(\ref{5566}). The development of the covariant 
derivative leads to the appearance of two terms, a free particle one and a 
fermion-boson coupling term,
\begin{equation}
{\cal L}=  \bar{ \psi} \; (i \gamma ^{\mu} \partial_{\mu} -m )\;  \psi - g 
\;  \bar{ \psi} \;  \gamma ^{\mu}  t_{a} W^{a}_{\mu}\;  \psi.
\end{equation}
Let us now consider an infinitesimal scale transformation of the 
fermion field,
\begin{equation}
\psi \rightarrow \psi \; e^{-ig\,  \delta \theta^b \,  t_b}.
\end{equation}
The requirement of the full Lagrangian invariance under this transformation 
involves also the coupling term. Let us consider the transformation of this 
term, except for the $W_{\mu}$ contribution,
\begin{equation}
 \bar{ \psi} \;  \gamma ^{\mu}  t_{a} \;  \psi \rightarrow
   \bar{ \psi} \; e^{ig\,  \delta \theta^b \,  t_b}  \gamma ^{\mu}  t_{a} \; 
 \psi\; e^{-ig\,  \delta \theta^b \,  t_b}.
\end{equation}
Accounting for the fact that this is an infinitesimal transformation, it 
becomes
\begin{eqnarray}
 \bar{ \psi} \; (1+ig\,  \delta \theta^b \,  t_b)  \gamma ^{\mu}  t_{a} \;  
\psi\; (1-ig\,  \delta \theta^b \,  t_b)
 = \bar{ \psi} \gamma ^{\mu}  t_{a} \;  \psi + i g \, \bar{ \psi} \gamma 
^{\mu} \; \delta \theta^b \; (t_b t_a -t_a t_b)\psi.
\end{eqnarray}
We replace the commutator $t_b t_a -t_a t_b$ by its expression in Eq.~(\ref{4011}), and we obtain
\begin{equation}
 \bar{ \psi} \;  \gamma ^{\mu}  t_{a} \;  \psi \rightarrow
  \bar{ \psi}\;  \gamma ^{\mu}  t_{a} \;  \psi - g \, \bar{ \psi}\;  \gamma 
^{\mu} \; \delta \theta^b \; f^{c}_{ba} t_c \;  \psi.
 \end{equation}
 Then the requirement of invariance could be fulfilled only provided 
the transformation of the field $W^a_{\mu}$ itself involves a new term (in 
addition to the Abelian term $ \partial_{\mu} \delta \theta^a$), i.e.,
 \begin{equation}
 W^a_{\mu} \rightarrow W^a_{\mu}  + \delta W^a_{\mu}.
\end{equation}
 The transformation of the full coupling term now reads
 \begin{equation}
 \bar{ \psi} \;  \gamma ^{\mu}  t_{a} W^{a}_{\mu}\;  \psi \rightarrow
  [( \bar{ \psi}\;  \gamma ^{\mu}  t_{a} \;  \psi) - g \, f^{c}_{ba} \; 
\delta \theta^b \;  ( \bar{ \psi}\;  \gamma ^{\mu}t_c \;  \psi)] \; 
[W^a_{\mu} + \delta W^a_{\mu}].
  \end{equation}
Neglecting the second order term in the elementary variations and using the 
fact that we can interchange the running indices, we see 
that this expression is invariant provided
   \begin{equation}
 \bar{ \psi} \;  \gamma ^{\mu} \;  \{ t_{a} [ \delta W^a_{\mu}  - g \, 
f^{a}_{bc} \; \delta \theta^b  \, W^c_{\mu}] \}\;  \psi=0.
  \end{equation} 
One general solution, independent of the $t_a$'s, to the requirement of the 
Lagrangian invariance in the non-Abelian case is therefore
  \begin{equation}
 \delta W^a_{\mu}=g \, f^{a}_{bc} \; \delta \theta^b  \, W^c_{\mu}.
  \end{equation} 
Finally, under an infinitesimal scale transformation $\delta \theta^b$, the 
non-Abelian gauge boson field $ W^a_{\mu}$ transforms as
 \begin{equation}
W^a_{\mu} \rightarrow W'^a_{\mu}= W^a_{\mu}+\partial_{\mu} \delta \theta^a
+g \, f^{a}_{bc} \; \delta \theta^b  \, W^c_{\mu}.
  \end{equation} 
We recognize here once again a standard transformation of non-Abelian
gauge theories, which is now derived from the basic transformations on the 
$\eta_a$'s of Eq.~(\ref{0027}).

We can finish as usual the development of standard Yang-Mills theory. 
The gauge field self-coupling term, $-{\frac{1}{4}}F_{\mu\nu }F^{\mu \nu }$, 
is retained as the simplest invariant scalar that can be
added to the Lagrangian. It is defined as follows.

First, one defines the Yang-Mills field,
\begin{equation}
A_{\mu} \equiv t_a W^a_{\mu},
\end{equation}
which yields the covariant derivative of Eq.~(\ref{5566}) under the standard 
form,
\begin{equation}
D_{\mu} = \partial_{\mu} + igA_{\mu}.
\end{equation}
Then, one establishes the analogue of the Faraday tensor of electromagnetism, 
by defining
\begin{equation}
F^a_{\mu \nu} \equiv \partial_{\mu} W^a_{\nu} - \partial_{\nu} W^a_{\mu} 
- g f^a_{bc} W^b_{\mu} W^c_{\nu}
\end{equation}
and
\begin{equation}
F_{\mu \nu} \equiv t_a F^a_{\mu \nu},
\end{equation}
which gives
\begin{equation}
F_{\mu \nu} = \partial_{\mu} A_{\nu} - \partial_{\nu} A_{\mu} + ig[A_{\mu}, 
A_{\nu}].
\end{equation}
One adds to the Lagrangian density ${\cal L}$ a kinetic term for the free 
Yang-Mills gauge field,
\begin{equation}
{\cal L}_A = -{\frac{1}{4}}F_{\mu\nu }F^{\mu \nu }.
\end{equation}
This form is justified by the same reasons as in the standard theory (namely, it must be a scalar and constructed from the fields and not from the potentials, which are gauge dependent).
The Euler-Lagrange equations therefore read
\begin{equation}
\partial _{\mu} F^{\mu \nu} + ig [A_{\mu}, F^{\mu \nu}] = 0.
\end{equation}
Introducing the Yang-Mills derivative operator,
\begin{equation}
\bigtriangledown_{\mu} = \partial_{\mu} + ig[A_{\mu}, \;\;],
\end{equation}
one finally obtains the standard Yang-Mills equations which generalize to 
the non-Abelian case the source-free Maxwell equation,
\begin{equation}
\bigtriangledown_{\mu} F^{\mu \nu} = 0.
\end{equation}
We are therefore provided with a fully consistent gauge theory obtained as a
consequence of scale symmetries issued from a geometric space-time
description.

\section{Conclusion}
\label{s.c}

%**************************************

In the present paper our purpose has been to give a physical meaning to the
various items entering the gauge field theories in the framework
of scale relativity, extending to the non-Abelian case the results of
previous works devoted to the understanding of the Abelian gauge-invariant
theory of electromagnetism.

We have so far reached an understanding of the nature of gauge 
transformations, in terms of a geometric space-time description. We decompose, for simplification purpose, the dynamics 
emerging from displacements in the fractal space-time of scale relativity 
into (i) transformations occuring on the scale variables in the framework 
of a nondirectly observable local ``scale space'' coupled to 
(ii) displacements in space-time. The scale variables become thus functions 
of the space-time coordinates.

The gauge charges appear as the generators of the set of
scale transformations applied to a generalized action, therefore
emerging from the scale symmetries of the dynamical fractal space-time.
Considering the transformation laws verified by the scale variables, we are
able to establish how the various physical quantities transform
under these laws and to recover the standard gauge theory form of these
transformations.

We are now provided with a theory where the gauge group is no more defined
through its only action on the physical objects, as in the standard framework,
but as the transformation group of the scale variables, and where the boson
fields and the charges are given a physical meaning. We have established the 
following correspondences between the standard gauge theory items and the 
scale relativistic tools:
\begin{itemize}
\item gauge transformations $\leftrightarrow$ scale transformations in scale-space,
\item internal gauge space $\leftrightarrow$ local ``scale space'',
\item gauge fields $\leftrightarrow$ manifestations of the fractal and scale-relativistic geometry of space-time (analogues of the Christoffel symbols issuing from 
the curvature of space-time in general relativity),
\item gauge charges $\leftrightarrow$ conservative quantities, conjugate to 
the scale variables, originating from the symmetries of the ``scale space'' 
and generators of the scale transformation group.
\end{itemize}
Since, in the present study, our aim was to recover the standard description of current gauge theory, we have, in the main part of the work, retained a general form for the scale variables.  However we have shown that, whatever will be their more specific form, the gauge set will contain in any case the U(1) $\times$ SU(2) electroweak theory group as subset. 

In future works, we shall study other sets of transformations that can be derived from the present study, where 
the scale variables will be given new precise definitions and which 
hopefully could yield hypercharge, color and maybe new developments in 
gauge field theory. We shall also consider in more details the issues of the fermion sectors, of the mass and charge renormalization and of the Higgs field.

{\it Acknowledgements}. The authors are grateful to Dr. J.E. Campagne for helpful 
remarks on an earlier version of the paper, and to a referee for his useful comments.

%***********

\end{document}